\documentclass[11pt]{article}
\usepackage{graphicx}
\title{\bf On the collision of two projectiles on two targets in the BFKL approach}
\author{M.A.Braun\\
Dep. of High Energy physics,
 Saint-Petersburg State University,\\
198504 S.Petersburg, Russia}

\newcommand\lra{\leftrightarrow}
\newcommand\beq{\begin{equation}}
\newcommand\eeq{\end{equation}}

\begin{document}

\maketitle

{\bf Abstract}\\
High-energy collisions of two nucleons on two nucleons are studied in the
BFKL approach in the leading approximation in $\alpha_sN_c$.
Diagrams with redistribution of colour are considered. It is found
that intermediate BKP states consisting of 4 reggeized gluons give a
contribution which may be leading in deuteron-deuteron scattering and thus
experimentally observable.

\section{Introduction}
Collisions of two heavy nuclei  have long occupied a prominent
part of the experimental and theoretical studies in strong
interactions at high energies. Unlike the case of DIS  theoretical
analysis of these processes turned out to be quite complicated.
The most advanced calculations have been made in the framework of
the Colour Glass Condensate approach, where they heavily rely on
numerical Monte- Carlo method on the lattice  to study evolution
of the classical gluonic field in the course of collisions
~\cite{krasnitz, nara, lappi}. In comparison, analytic methods
applied to coliision of heavy nuclei have only given modest and
approximate results ~\cite{kovchegov,balitski, dusling}. So to
understand the problem it seems natural to start not from this
general case, but from the simplest generalization of the
well-know results for DIS to the collision of two nucleons with
two nucleons. The immediate physical application is of course
the case of deuteron-deuteron collisions.
An alternative view is to consider this as a particular
contribution to the amplitude for the collision of two heavy
nuclei A and B, which generally contains contributions from collisions of
any number of  nucleons in the projectile nucleus on any number of
nucleons in the target.
For large atomic numbers
$A$ and $B$ the scattering amplitude is effectively unitarized by
eikonalization of its connected part $E(b)$ of the corresponding
diagrams:
\beq {\cal A}_{AB}=2is\int d^2b\Big(1-e^{-E(b)}),
\label{eik} \eeq
where $b$ is the impact parameter of the
collision and $s$ is the c.m energy squared for a pair of
colliding nucleons. Our study then refers to the part of the
eikonal function coming from the collision of two nucleons from the
projectile nucleus on two nucleons from the target nucleus.
Due to eikonalization  the total nucleus-nucleus cross-section for heavy nuclei
 does not practically depend on a concrete value of the eikonal function.
The latter
is then large and  its changes even by several times do not mean much for the
total cross-section, which remains essentially geometrical. However even for heavy
nuclei it may have influence on some
specific processes, like exclusive rare production, where it may change
the absorptive factor, and of course on the inclusive cros-sections.

\begin{figure}
\hspace*{2.5 cm}
\begin{center}
\includegraphics[scale=0.65]{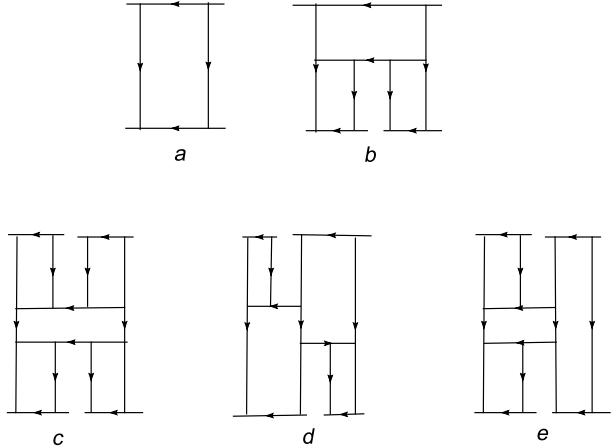}
\end{center}
\caption{Lowest order diagrams for the eikonal: two gluon exchange (a) and with
transitions to 2 (b) or 3 (c,d) intermediate gluons}
\label{fig0}
\end{figure}

It is instructive to see relative orders of magnitude
for various contributions to the eikonal function $E(b)$.
Naturally the leading order in both coupling constant $\alpha_s$
and number of colours $N_c$ is given by the double gluon exchange,
Fig. \ref{fig0} $a$. For simplicity, instead of nucleons we shall
consider quarks as elementary scattering centers in the nuclei,
invoking the colour neutrality by projecting onto the colourless
$t$-channels.

\begin{figure}
\hspace*{2.5 cm}
\begin{center}
\includegraphics[scale=0.5]{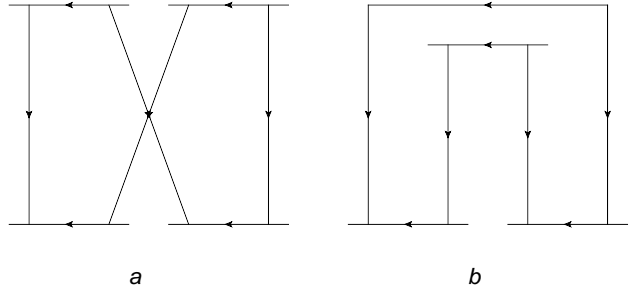}
\end{center}
\caption{Lowest order diagram with a redisiribution of color
(diagrams with crossing gluon lines should be added)}
\label{fig1}
\end{figure}

Note that the total diagram for 2 by 2 scattering involves also the
product of two such diagrams. However the eikonal function will include
only one of them. If we forget about  couplings inside the scattering
centers, its contribution does not depend on $\alpha_s$ and is
proportional to $N_c^2$.
Other connected diagrams may be classified by the minimal number of
exchanged gluons $n_{min}$. We shall discuss their order of magnitude
related
to the double gluon exchange, Fig. \ref{fig0} $a$.
The diagram in Fig. \ref{fig0} $b$ with $n_{min}=2$ has this relative order
$\alpha_s^2N_c^2$. The diagram in Fig.\ref{fig0},$c$ with $n_{min}=2$
has the relative order $\alpha_s^4N_c^4$. Diagrams $d$ and $e$ with $n_{min}=3$
have order $\alpha_s^3 N_c^3$. Finally typical diagrams with $n_{min}=4$
and redistribution
of colour, which makes them connected, are shown in Fig. \ref{fig1}
and have the same order as Fig. \ref{fig0}$a$, so that their relative
order is unity.
To finally estimate the  weights of these diagrams one has to take into
account
that diagrams with $n_{min}=2$ and 3 involve one or two intermediate
rapidities at which the
initial and final 4 gluons fuse into the intermediate ones and so are
proportional to  $y$ or $y^2$ where $y$ is the overall rapidity. So their
final order will be $\alpha_s^2N_c^2y$, $\alpha_s^4N_c^4y^2$ and
$\alpha_s^3 N_c^3y^2$.
In the BFKL kinematics one assumes $\alpha_sN_cy\sim 1$, $\alpha_sN_c<<1$
so that orders of diagrams Fig. \ref{fig0}$b$, $c$, $d$ and $e$
 become $\alpha_sN_c$, $\alpha_s^2N_c^2$ and $ \alpha_sN_c$
respectively.
This shows that apart from the double gluon exchange the dominant
contribution comes from the diagrams with colour redisitribution,
Fig. \ref{fig1}.
Of course this result has been obtained in the lowest order,
but it remains valid also in
higher orders when the leading order will be multiplied by
powers of $\alpha_sN_cy\sim 1$.

In this study we shall consider the connected part of
the forward scattering
ampitude of two projectile centers (quarks) on two target centers
(also quarks).
As we have demonstrated, in the lowest order
it is just the simple rearrangement amplitude shown in
Fig. \ref{fig1} in two different forms: one (a) symmetric in projectiles and targets
and another (b) showing  the intermediate states in the $s$-channel. In fact both in Fig. \ref{fig1}
$a$ and $b$ one should
also take into account all diagrams with crossed vertical lines
(16 diagrams in all).
We would like to study all possible inclusions of interactions between the gluons.
They are realized by the BFKL interactions $V_{ij}$ between gluond $i$ and
$j$ ~\cite{lipatov}.
We shall work in the leading approximation in the number of colours $N_c$.
As compared to the disconnected diagram with two
pomeron exchanges of the leading order in $N_c$,
all contributions studied in the following will be of the order $1/N_c^2$
and so subdominant
in the large $N_c$ limit. However in the eikonal function they will contain
an extra nuclear factor
of the relative order $A^{2/3}$ for collisions of two nuclei of the same
atomic number $A$ and so in
fact may be of the same order or even greater than the leading
contribution in $N_c$.

\begin{figure}
\hspace*{2.5 cm}
\begin{center}
\includegraphics[scale=0.5]{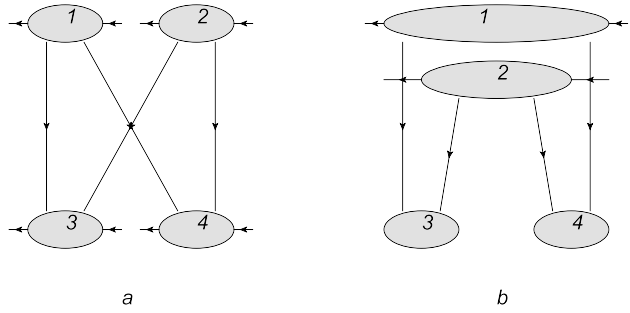}
\end{center}
\caption{Diagrams for the direct transition of pomerons with
the redistribution of colour}
\label{fig2}
\end{figure}

We can separate all contributions into three classes.
First, interactions may occur only between gluons attached to the same
projectile
or target and so being in the vacuum state. Such interactions will convert
pairs
of gluons attached to the projectiles and targets in Fig. \ref{fig1}
together with their crossing into fully developed pomeron
Green functions, so that the quarks representing our projectiles and
targets will convert
to pomerons, see Fig. \ref{fig2}. For heavy nuclei reggeized gluon
splitting will further
convert simple pomerons into fan diagrams made of pomerons, which are summed into the solution
of the Balitski-Kovchegov (BK) equation, BK wave functions
~\cite{balitski1, kovchegov1}.

\begin{figure}
\hspace*{2.5 cm}
\begin{center}
\includegraphics[scale=0.5]{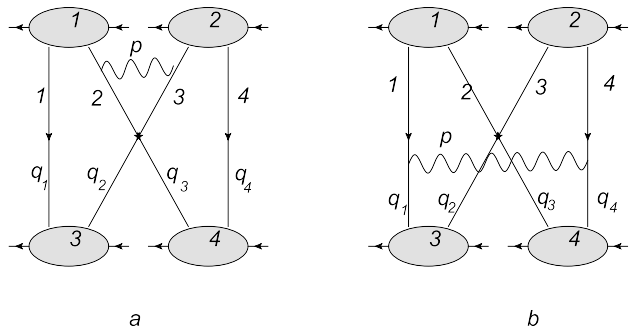}
\end{center}
\caption{Diagrams with the redistribution of colour and one interaction beiween
the pomerons of the projectile and target}
\label{fig3}
\end{figure}

Second class of the diagrams are formed form those of the first class
with one interaction
between the gluons $V_{23}$ or $V_{14}$, which does not connect any pair of gluons in the
vacuum state (Fig. \ref{fig3}).

\begin{figure}
\hspace*{2.5 cm}
\begin{center}
\includegraphics[scale=0.85]{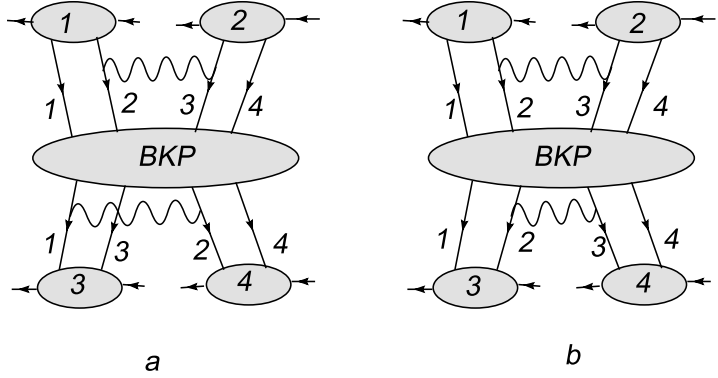}
\end{center}
\caption{Diagrams with the intermediate BKP states}
\label{fig4}
\end{figure}

Third class of the diagrams are those where from each side there
appear interactions which
connect vacuum pairs of gluons with the so-called BKP states ~\cite{bartels, kwie} between them.
(Fig. \ref{fig4}).

Before actually calculating all these contributions, in Section 2 we develop a
multirapidity formalism which makes it easier to sew four pomerond
into a single amplitude.

Let the momenta of the nucleons in the projectile be $k$ and those in the
target be $l$. In the c.m. system of a pair of nucleons from the
projectile and target we have
\[ k_+=l_-,\ \ k_-= k_\perp=l_+=l_\perp=0,\ \ s=2k_+l_-\]
The amplitude for nucleus-nucleus scattering can be separated into its
high-energy part $H$ and two nuclear factors (see Appendix 1.).
The imaginary part of  $H$ in its turn can be presented in the form
\beq
{\rm Im}\,H=-(2\pi)^2\delta(\kappa_+)\delta(q_-)4s^2N_c^2 D,
\label{ddef}
\eeq
where $\kappa$ and $q$ are the momenta transferred ito the
projectile and target nuclei respectively with $\kappa_-=\kappa_\perp=
q_+=q_\perp=0$ and $D$ is a real function which is the sum of connected
diagrams for the forward scattering of two-nucleons on two nucleons written
as integrals in rapidity and transverse momentum space.
At fixed impact parameter $b$ the  cross-section for the part of nucleus-nucleus scattering
coming from interactions of two nucleons from the projectile on two nucleons from
target is related to $D$ as
\beq
\sigma_{AB}^{(2)}(b)=-2N_c^2D\int d^2b_AT_A^2(b_A)T_B^2(b-b_A)
\label{sigab}
\eeq
and for deuteron-deuteron scattering
\beq
\sigma_{dd}=-2N_c^2D\Big<\frac{1}{2\pi r^2}\Big>_d^2
\label{sigdd}
\eeq
(see Appendix 1.).

Note in conclusion that the contribution of simplest diagrams with colour rearrangements have
been considered in literature in relation to correlations between a pair of produced gluons
~\cite{dumitru}.
We postpone discussion of the inclusive gluon production to future
publications, since in our BFKL-Bartels approach it involves the study of
possible cuts of the forward scattering amplitudes, which
 for 2 by 2 scattering requires special attention  to the well-known AGK
 cancellations.
Meanwhile we have to stress that already on the level of the total
cross-section,  apart from the
pomeron (or BK) wave function, our amplitudes  involve more complicated
objects made of 4 reggeized gluons (the BKP states).
In the Colour Glass Condensate approach similar
complicated structures begin to appear only on the level of double
inclusive cross-sections.

\section{Pomeron in the multirapidity formalism}

\def\ep{\epsilon}

In this section we introduce a formalism in which each pomeron has its own rapidity,
which corresponds to the standard Feynman diagram
technique. In this formalism construction of amplitudes with colour rearrangment becomes
much simpler.
Recall that in the BFKL approach the amplitude ("wave function") $P(y)$ at a given rapidity
$y$ is obtained from its value at $y=0$ by the transformation
\beq
P(y)=e^{-yH}P(0).
\label{eq1}
\eeq
Here the BFKL Hamiltonian is
$
H=H^{(0)}+V.
$
The unperturbed Hamiltonian $H^{(0)}$ is a sum of Regge trajectories
$\omega(k)$ with a minus sign.
For gluons 1 and 2
$
H^{(0)}_{12}=-\omega_1-\omega_2
\label{eq3}
$
or, in the momentum representation,
\beq
<k'_1,k'_2|H^{(0)}|k_1,k_2>=-(2\pi)^4\delta^2(k_1-k'_1)\delta^2(k_2-k'_2)\Big(\omega(k_1)+\omega(k_2)\Big).
\label{eq4}
\eeq
The potential energy $V$ is given by the pair BFKL interaction. Symmetrizing in initial and final
reggeons we have
\beq
V_{12}=(T_1T_2)(2\pi)^2\delta^2(k_1+k_2-k'_1-k'_2)
v(k'_1,k_2|k_1,k_2),
\label{eq5a}
\eeq
where
\beq
v(k'_1,k'_2|k_1,k_2)=\frac{g^2}{2\pi k_1k_2k'_1k'_2}
\Big(\frac{k_1^2{k'_2}^2+k_2^2{k'_1}^2}{(k_1-k'_1)^2}-(k_1+k_2)^2\Big)
\label{eq5}
\eeq
and $T_1$ and $T_2$ are  colours of the two gluons.

Evolution law (\ref{eq1}) mimics the standard evolution in time provided one makes the
substitution
\beq
t\to -iy,\ \ y\to it,\ \ y>0.
\label{ty}
\eeq
By definition $P(y)=0$ at $y<0$.
We shall obtain our diagrammatic technique studying  evolution in time. Evolution
in rapidity will be obtained making the analytic continuation (\ref{ty}) at $y>0$ in  final formulas.

Passing to the multirapidity formalism we shall additionally characterize each reggeon by its "energy"
$\epsilon$. We shall take the reggeon propagator as a function of $\ep$ in the form
\beq
\Delta(\ep,k)=\frac{i}{\ep+\omega(k)+i0}.
\eeq
Then as a function of time
\beq
\Delta(t,k)=i\int \frac{d\ep}{2\pi }e^{-i\ep t}
\frac{1}{\ep+\omega(k)+i0}=\theta(t)e^{it\omega(k)}.
\eeq
Taking $it\to y$ at $t>0$ and $y>0$ we get the desired reggeon propagator
\beq
\Delta(y,k)=\theta(y)e^{y\omega(k)}.
\eeq

The interaction between reggeons will be given by (\ref{eq5a}) with factor
$(-i)$ and
with an additional factor responsible  for  conservation of the total energy
\beq
\hat{V}_{12}=-i(T_1T_2)(2\pi)^3\delta(\ep_1+\ep_2-\ep'_1-\ep'_2)\delta^2(k_1+k_2-k'_1-k'_2)
v(k'_1,k_2|k_1,k_2).
\label{eq5a1}
\eeq

These Feynman rules allow to construct amplitudes for the interaction of any number of
reggeons provided this number does not change.

Our first task is to check that for a pair of reggeons the sum of all thus consructed Feynman diagrams
is equivalent to the standard BFKL equation.

\begin{figure}
\hspace*{2.5 cm}
\begin{center}
\includegraphics[scale=0.5]{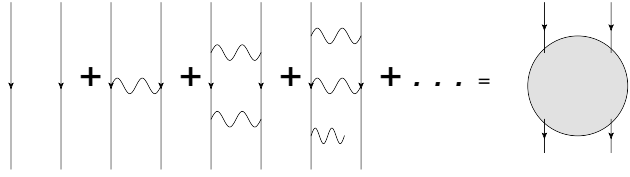}
\end{center}
\caption{The pomeron Green function ${\cal G}$}
\label{fig5}
\end{figure}

\begin{figure}
\hspace*{2.5 cm}
\begin{center}
\includegraphics[scale=0.5]{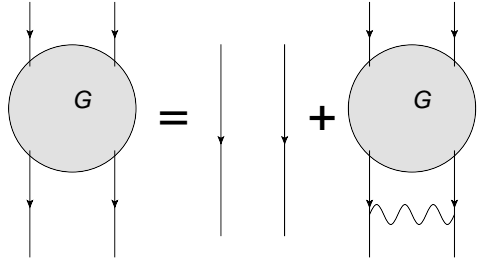}
\end{center}
\caption{Equation for the pomeron Green function ${\cal G}$}
\label{fig6}
\end{figure}

The Green function for a pair of interacting reggeons
${\cal G}(\ep'_1k'_1,\ep'_2k_2|\ep_1k_1,\ep_2k_2)$, which is
illustrated in Fig. \ref{fig5},
obeys an
equation which is graphically shown in Fig. \ref{fig6}.

Separating from ${\cal G}$ the $\delta$-functions which corresponds to
energy-momentum conservation
\beq
{\cal G}(\ep'_1k'_1,\ep'_2k_2|\ep_1k_1,\ep_2k_2)=(2\pi)^3
\delta(\ep'_1+\ep'_2-\ep_1-\ep_2)\delta^2(k'_1+k'_2-k_1-k_2)
{\cal G}_{EK}(\ep'_1k'_1|\ep_1k_1),
\label{dred}
\eeq
where $E=\ep_1+\ep_2=\ep'_1+\ep'_2$ and $K=k_1+k_2=k'_1+k'_2$.
we find a  Bethe-Salpeter type equation for ${\cal G}_{EK}$
\beq
{\cal G}_{EK}(\ep'_1k'_1|\ep_1k_1)=(2\pi)^3\delta(\ep'_1-\ep_1)
\delta^2(k'_1-k_1)\Delta_1\Delta_2
+
\Delta'_1\Delta'_2
\int \frac{d\ep''_1d^2k''_1}{(2\pi)^3}
\hat{V}(k'_1,k'_2|k''_1,k''_2){\cal G}_{EK}(\ep''_1k''_1|\ep_1k_1),
\label{dequ}
\eeq
where we denote $\Delta_1=i/(\ep_1+\omega_1)$, $\Delta'_1=i/(\ep'_1+\omega'_1)$
and so on and all $\omega$'s are assumed to have a small positive imaginary part.

In this equation the interaction is energy independent, as in theories with non-relativistic potentials.
So it can be easily transformed into a Schroedinger-like equation.
Indeed if we present
\beq
{\cal G}_{EK}(\ep'_1k'_1|\ep_1k_1)=(2\pi)^3\delta(\ep'_1-\ep_1)
\delta^2(k'_1-k_1)\Delta_1\Delta_2
+\Delta'_1\Delta'_2
\hat{T}_{EK}\Delta_1\Delta_2
\label{tdef}
\eeq
then $\hat{T}_{EK}(\ep'_1k'_1|\ep_1k_1)$ does not depend on initial nor
final energies $\ep_1$, $\ep_2$, $\ep'_1$ and $\ep'_2$.
The equation for $\hat{T}$ takes the form
\beq
\hat{T}_{EK}(k'_1|k_1)=\hat{V}(k'_1,k'_2|k_1,k_2)+
\int
i\frac{d^2k''_1}{(2\pi)^2}
\frac{\hat{V}(k'_1,k'_2|k''_1,k''_2)\hat{T}_{EK}(k''_1|k_1)}{E+\omega''_1+\omega''_2+i0}
\label{teq1}
\eeq
Recalling that $\hat{V}=-iV$, where $V=(T_1T_2)v$, and putting
$\hat{T}=-iT$
we rewrite Eq. (\ref{teq1})
in the operatorial form
\beq
T_{EK}=V+VR_{EK}T_{EK}
\label{teq2}
\eeq
where
\beq
R_{EK}=(E-H^{(0)}+i0)^{-1}
\label{rdef1}
\eeq
is the resolvent of the unperturbed Hamiltonian.
Eq. (\ref{teq2}) is the standard equation for the $T$-matrix.
The standard BFKL Green function is defined by $T_{EK}$ as
\beq
G_{EK}=R_{EK}+R_{EK}T_{EK}R_{EK}=(E-H)^{-1}
\label{gviat}
\eeq
The Bethe-Salpeter
Green function ${\cal G}$ is expressed via the Schroedinger one $G$,
as
\[
{\cal G}_{EK}(\ep'_1k'_1|\ep_1k_1)=(2\pi)^3\delta(\ep'_1-\ep_1)
\delta^2(k'_1-k_1)\Delta_1\Delta_2
+i
(2\pi)^2\delta^2(k'_1-k_1)[E+\omega_1+\omega_2]
\Delta'_1\Delta'_2\Delta_1\Delta_2
\]\beq
-i
\Delta'_1\Delta'_2[E+\omega'_1+\omega'_2]G_{EK}(k'_1|k_1)
[E+\omega_1+\omega_2]\Delta_1\Delta_2.
\label{dviag}
\eeq

\begin{figure}
\hspace*{2.5 cm}
\begin{center}
\includegraphics[scale=0.5]{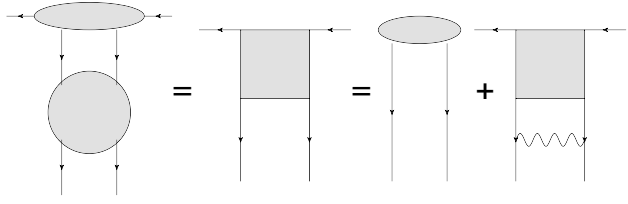}
\end{center}
\caption{Equation for the pomeron wave function ${\cal P}$}
\label{fig7}
\end{figure}

Now we pass to the BFKL function proper. Integrating the Green function
${\cal G}_{EK}(\ep'_1k'_1|\ep_1k_1)$ with the impact factor
$\rho_{EK}(k'_1)/(k'_1k'_2)$
we obtain a function which describes the pomeron in the multirapidity
formalism
\beq
{\cal P}_{EK}(\ep_1k_1)=\int \frac{d\ep'_1d^2k'_1}{(2\pi)^3k'_1k'_2}\rho_{EK}(k'_1){\cal G}_{EK}(\ep'_1k'_1|\ep_1k_1).
\eeq
The equation for it easily follows from Fig. \ref{fig7}
\beq
{\cal P}_{EK}(\ep_1k_1)={\cal P}_{EK}^{(0)}(\ep_1k_1)
-i
\Delta_1\Delta_2
\int \frac{d\ep'_1d^2k'_1}{(2\pi)^3}
V(k_1,k_2|k'_1,k'_2){\cal P}_{EK}(\ep'_1k'_1),
\label{deq}
\eeq
where $\ep_2=E-\ep_1$, $k_2=K-k_1$ and $\omega_2=\omega(k_2)$.
If we introduce the "amputated" wave function by
\beq
{\cal P}_{EK}(\ep_1k_1)=
F_{EK}(\ep_1,k_1)\Delta_1\Delta_2
\eeq
then $F$ does not depend on energy $\ep_1$
and satifies
\beq
F_{EK}(k_1)=F_{EK}^{(0)}(k_1)
+
\int \frac{d^2k'_1}{(2\pi)^2}
\frac{V(k_1,k_2|k'_1,k'_2)F_{EK}(\ep'_1k'_1)}
{E+\omega'_1+\omega'_2}.
\label{tfeq}
\eeq
So if we define a new wave function
\beq
P_{EK}(k_1)=\frac{F_{EK}(k_1)}
{E+\omega_1+\omega_2}
\label{defp}
\eeq
it will obey the equation
\beq
(E+\omega_1+\omega_2)P_{EK}(k_1)=F_{EK}^{(0)}(k_1)+
\int \frac{d^2k'_1}{(2\pi)^2}
V(k_1,k_2|k'_1,k'_2)P_{EK}(k'_1),
\label{peq}
\eeq
which is the standard BFKL equation (for the pomeron or gluon
depending on the value of $(T_1T_2)$ in $V$).

Note that BS function ${\cal P}$ turns out to be related to $P$ as
\beq
{\cal P}_{EK}(\ep_1k_1)=
\frac{E+\omega_1+\omega_2}
{(\ep_1+\omega_1)(\ep_2+\omega_2)}P_{EK}(k_1).
\eeq
It depends on the individual energies.
It is remarkable that although for the pomeron $P$ is infrared safe, the
coresponding BS function ${\cal P}$ does not look so.

\section{Four pomerons sewed with the redistribution of colour}
Now we pass to the simplest amplitude $D_1$ in which two pomerons from
the projectile (reggeon pairs (1,2) and (3,4)) are directly coupled to two pomerons in the target
 (reggeon pairs (13) and (24)) illustrated in Fig. \ref{fig2}. All pomerons will have their total momenta equal to zero,
which will be tacitly assumed in the following. The pomerons from the
projectile have their total energies $E_{12}$ and $E_{34}$ and
those from the target energies $E_{13}$ and $E_{24}$.
Energy conservation requires
$
E_{12}+E_{34}=E_{13}+E_{24}.
$
Correspondingly the amplitude
$D_1(E_{12},E_{34},E_{13},E_{24})$ will include factor
$ 2\pi \delta(E_{12}+E_{34}-E_{13}-E_{24})$
which will be suppressed in the following.
As seen from Fig. \ref{fig2}, the internal integrations will include a
single transverse momentum $q$ and a single pomeron energy, say, $\ep_1$.
The remaining energies will be expressed via $\ep_1$ as follows
\beq
\ep_2=E_{12}-\ep_1,\ \ \ep_3=E_{13}-\ep_1,\ \
\ep_4=E_{24}-\ep_2=E_{24}-E_{12}+\ep_1.
\label{eps}
\eeq
In terms of the amputated pomeron function $F$ the amplitude will be given by
\beq
D_1(E_{12},E_{34},E_{13},E_{24})=
\int\frac{d\ep_1 d^2q}{(2\pi)^3}
\frac{F_{E_{12}}(q)F_{E_{34}}(q)F_{E_{13}}(q) F_{E_{24}}(q)}
{(\ep_1+\omega(q))(\ep_2+\omega(q))(\ep_3+\omega(q))(\ep_4+\omega(q))},
\label{ampli}
\eeq
where all $\omega$'s are assumed to have a small positive imaginary part.

Since $F$'s do not depend on $\ep_1$ we can do integration on $\ep_1$ explicitly. We have an integral
\[
I=\int\frac{d\ep_1}{2\pi}
\frac{1}{\ep_1+\omega(q)+i0}\,
\frac{1}{E_{12}-\ep_1+\omega(q)+i0}\,
\frac{1}{E_{13}-\ep_1+\omega(q)+i0}\,
\frac{1}{E_{24}-E_{12}+\ep_1+\omega(q)+i0}.
\]\beq
=
-\frac{i}{E_{13}-E_{12}}\,
\Big\{
\frac{1}{E_{12}+2\omega(q)+i0}\,
\frac{1}{E_{24}+2\omega(q)+i0}
-
\frac{1}{E_{13}+2\omega(q)+i0}\,
\frac{1}{E_{34}+2\omega(q)+i0}\Big\}.
\label{iint}
\eeq

Now we express our amputated pomerons $F$ via full ones $P$ using
Eq. (\ref{defp}). In terms of  $P$ the amplitude will be given by
\beq
D_1(E_{12},E_{34},E_{13},E_{24})=
-i\int\frac{ d^2q}{(2\pi)^2}
\Big(E_{12}+E_{34}+4\omega(q)\Big)
P_{E_{12}}(q)P_{E_{34}}(q)P_{E_{13}}(q) P_{E_{24}}(q).
\label{ampli1}
\eeq
Immediately the question of its infrared safeness arises,
since the integrand contains the reggeon trajectory $\omega(q)$.

Next we  study  evolution in rapidity.
\[
D_1(y)=
\int \frac{dE_{12}dE_{34}dE_{13}dE_{24}}{(2\pi)^4}
2\pi\delta(E_{12}+E_{34}-E_{13}-E_{24})e^{-y(E_{12}+E_{34}}D_1(
E_{12},E_{34},E_{13},E_{24}=\]\[
-i\int\frac{ d^2q}{(2\pi)^2}\Big(4\omega(q)-\frac{\partial}{\partial y}\Big)
\int \frac{dE_{12}dE_{34}dE_{13}dE_{24}}{(2\pi)^4}
2\pi\delta(E_{12}+E_{34}-E_{13}-E_{24})
\]\beq
e^{-y(E_{12}+E_{34})}
P_{E_{12}}(q)P_{E_{34}}(q)P_{E_{13}}(q) P_{E_{24}}(q).
\label{ampli2}
\eeq
We first consider
evolution of the integrand of the momentum integral  in time:
\[
Z(t)=\int \frac{dE_{12}dE_{34}dE_{13}dE_{24}}{(2\pi)^4}
2\pi\delta(E_{12}+E_{34}-E_{13}-E_{24})e^{-it(E_{12}+E_{34})}
P_{E_{12}}(q)P_{E_{34}}(q)P_{E_{13}}(q) P_{E_{24}}(q)\]\beq=
\int \frac{dE_{12}dE_{34}dE_{13}dE_{24}}{(2\pi)^4}
d\tau e^{i\tau(E_{12}+E_{34}-E_{13}-E_{24})}
e^{-it(E_{12}+E_{34})}
P_{E_{12}}(q)P_{E_{34}}(q)P_{E_{13}}(q) P_{E_{24}}(q).
\label{zdef}
\eeq
We use
\beq
\int dEP_E(q)e^{-itE}=\theta(t)P(t,q)
\label{pt}
\eeq
to obtain
\beq
Z(t)=\int d\tau \theta(t-\tau)P^2(t-\tau,q)\theta(\tau)P^2(\tau,q)=\theta(t)
\int_0^t d\tau P^2(t-\tau,q)P^2(\tau,q).
\eeq
Analytic continuation to rapidity gives
\beq
Z(y)=-i\int dy' \theta(y-y')P^2(y-y',q)\theta(y')P^2(y',q)=-i\theta(y)
\int_0^y dy' P^2(y-y',q)P^2(y',q),
\label{zy}
\eeq
so that finally
\beq
D_1(y)=-\theta(y)\int\frac{ d^2q}{(2\pi)^2}
\Big(4\omega(q)-\frac{\partial}{\partial y}\Big)
\int_0^y dy' P^2(y-y',q)P^2(y',q).
\label{ampli3}
\eeq

To check the correctness of the transition from time to rapidity
we study a simple model, in which $P_E$ is given by a BFKL pole
\beq
P_E=\frac{1}{E+a+i0},
\eeq
so that $P(y)=\theta(y)e^{ay}$.
In this case it is easy to find evolution of $Z(y)$  in rapidity by
explicit analytic continuation.
Let  for complex $z$
\beq
Z(z)
=\int \frac{dE dE_{12}dE_{13}}{(2\pi)^3}
e^{-zE}
P_{E_{12}}(q)P_{E-E_{12}}(q)P_{E_{13}}(q) P_{E-E_{13}}(q).
\eeq
We have
\[
\int\frac{dE'}{2\pi}P_{E'}P_{E-E'}=
\int\frac{dE'}{2\pi}\frac{1}{(E'+a+i0)(E-E'+a+i0)}=
-i\frac{1}{E+2a+i0}.\]
So we find
\beq
Z(z)=-\int\frac{dE}{2\pi}e^{-zE}\frac{1}{(E+2a+i0)^2}
=i\frac{\partial}{\partial E}e^{-zE}|_{E=-2a}=-ize^{2az}.
\label{eq55}
\eeq
But this integral only exists for pure imaginary $z=it$ when it is given
\beq
Z(t)=-\int\frac{dE}{2\pi}e^{-itE}\frac{1}{(E+2a+i0)^2}
=
+i\theta(t)\frac{\partial}{\partial E}e^{-itE}|_{E=-2a}=te^{2ait}\to -ize^{2az}.
\eeq

On the other hand, from (\ref{zy}) we find
\beq
Z(y)=-i\theta(y)
\int_0^y dy' e^{2a(y-y')}e^{2ay'}=-iye^{2ay}
\eeq
in accordance with (\ref{eq55}).
This confirms factor $(-i)$  in (\ref{zy}) and  so (\ref{ampli3}).

\section{Single interaction between the projectile and target pomerons}
Now we consider diagrams with a single interaction between the projectile and target
which cannot be included into the pomerons, that is $V_{23}$ or $V_{14}$, see
Figs. \ref{fig3}$a$ and $b$.

We start with the diagram shown in Fig. \ref{fig3}$a$. As before we suppress
the energy conservation factor $2\pi\delta(E_{12}+E_{34}-E_{13}-E_{24})$.
The diagram contains two loops and so integrations over $\ep_1,\ep_4, q_1$
 and $q_4$.
Energy-momenta of the gluons 1,2,3 and 4 before the interaction are
\beq
(\ep_1,q_1),\ \ (E_{12}-\ep_1,-q_1),\ \ (E_{34}-\ep_4,-q_4),\ \  (\ep_4,q_4).
\label{qini}
\eeq
After the interaction gluon 2 and 3 have their energy-momenta
$(E_{24}-\ep_4)$ and $(E_{31}-\ep_1,-q_1)$ respectively.
In terms of amputated pomerons $F$ the contribution from Fig. \ref{fig3} $a$ is
\beq
D_{2a}=i\int\frac{d\ep_1d\ep_4d^2q_1d^2q_4
F_{E_{12}}(q_1)F_{E_{34}}(q_4)F_{E_{13}}(q_1)F_{E_{24}}(q_4)V(-q_1,-q_4|-q_4,-q_1)}
{(2\pi)^6
(\ep_1+\omega_1)(E_{12}-\ep_1+\omega_1)(E_{13}-\ep_1+\omega_1)
(\ep_4+\omega_4)(E_{34}-\ep_4+\omega_4)(E_{24}-\ep_4+\omega_4)}.
\eeq
The prefactor includes  $-i$ from
the definition of $\hat{V}$ and $(-1)$ from 6 propagators.

Integration over energies factorizes into two integrals:
\beq
I_1=\int \frac{d\ep_1}{2\pi
(\ep_1+\omega_1+i0)(E_{12}-\ep_1+\omega_1+i0)(E_{13}-\ep_1+\omega_1+i0)}
=-i\frac{1}{(E_{12}+2\omega_1)(E_{13}+2\omega_1)}
\eeq
and a similar integral over $\ep_4$ which gives
\beq
I_2=-i\frac{1}{(E_{34}+2\omega_4)(E_{24}+2\omega_4)}.
\eeq
So we get
\beq
D_{2a}=-i\int\frac{d^2q_1d^2q_4
F_{E_{12}}(q_1)F_{E_{34}}(q_4)F_{E_{13}}(q_1)F_{E_{24}}(q_4)V(-q_1,-q_4|-q_4,-q_1)}
{(2\pi)^4(E_{12}+2\omega_1)(E_{13}+2\omega_1)
(E_{34}+2\omega_4)(E_{24}+2\omega_4)}.
\eeq

Recalling relation (\ref{defp}) between the amputated wave function $F$ and
pomeron we rewrite this as
\beq
D_{2a}=-i\int\frac{d^2q_1d^2q_4}{(2\pi)^4}
P_{E_{12}}(q_1)P_{E_{34}}(q_4)P_{E_{13}}(q_1)P_{E_{24}}(q_4)
V(-q_1,-q_4|-q_4,-q_1).
\label{aa}
\eeq

Now we pass to the amplitude corresponding to the diagram in Fig.
\ref{fig3} $b$.
Before the interaction gluons 1,2,3 and 4 have the same energy-momenta as before
(Eq(\ref{qini})). However now the momenta of gluons 1 and 4 change after the interaction
and become $(E_{13}-E_{34}+\ep_4,q_4)$ and $(E_{24}-E_{12}+\ep_1)$ respectively.

In terms of $F$ the amplitude is
\[
D_{2b}=i\frac{1}{(2\pi)^6}\times \]\beq
\int\frac{d\ep_1d\ep_4d^2q_1d^2q_4
F_{E_{12}}(q_1)F_{E_{34}}(q_4)F_{E_{13}}(q_4)F_{E_{24}}(q_1)V(q_1,q_4|q_4,q_1)}
{
(\ep_1+\omega_1)(E_{12}-\ep_1+\omega_1)(E_{24}-E_{12}+\ep_1+\omega_1)
(\ep_4+\omega_4)(E_{34}-\ep_4+\omega_4)(E_{13}-E_{34}+\ep_4+\omega_4)}.
\eeq

Again the integrals over energies factorize into two ones:
\beq
I_3=\int\frac{d\ep_1}{2\pi
(\ep_1+\omega_1+i0)(E_{12}-\ep_1+\omega_1+i0)(E_{24}-E_{12}+\ep_1+\omega_1+i0)}
=i\frac{1}{(E_{12}+2\omega_1)(E_{24}+2\omega_1)}
\eeq
and a similar integral over $\ep_4$
\beq
I_4=i\frac{1}{(E_{34}+2\omega_4)(E_{13}+2\omega_4)}.
\eeq

We get
\beq
D_{2b}=-i\int\frac{d^2q_1d^2q_4
F_{E_{12}}(q_1)F_{E_{34}}(q_4)F_{E_{13}}(q_4)F_{E_{24}}(q_1)V(q_1,q_4|q_4,q_1)}
{(2\pi)^4(E_{12}+2\omega_1)(E_{13}+2\omega_4)(E_{34}+2\omega_4)
(E_{24}+2\omega_1)}
\eeq
or in terms of pomerons $P$
\beq
D_{2b}=-i\int\frac{d^2q_1d^2q_4}{(2\pi)^4}
P_{E_{12}}(q_1)P_{E_{34}}(q_4)P_{E_{13}}(q_4)P_{E_{24}}(q_1)
V(q_1,q_4|q_4,q_1).
\label{ab}
\eeq

Let us separate the infrared stable and divergent parts in $D_2$.
For a pair of gluons 1 and 2 the BFKL Hamiltonian is
\[H=-\omega_1-\omega_2+(T_1T_2)v_{12},\]
where $v_{12}$ is given by (\ref{eq5}). In the vacuum $t$-channel $(T_1T_2)=-N_c$, so that the
infrared stable pomeron Hamiltonial is
\beq
H_P=-\omega_1-\omega_2-N_cv_{12}.
\label{hpom}
\eeq

In $D_2$  interaction connects different colour configurations with
gluons from the
projectile forming colourless pairs (1,2) and (3,4) and from the target
forming colorless pairs (1,3) and (2,4).
The transition matrix element of $(T_2T_3)$ entering the interaction is
$+N_c$ (see Appendix 2.).
So in terms of $v$ amplitude $D_{2a}$ is given by
\beq
D_{2a}=-i\int\frac{d^2q_1d^2q_4}{(2\pi)^4}
P_{E_{12}}(q_1)P_{E_{34}}(q_4)P_{E_{13}}(q_1)P_{E_{24}}(q_4)N_c
v(-q_1,-q_4|-q_4,-q_1).
\eeq
We present
\beq
N_cv(-q_1,-q_4|-q_4,-q_1)=
-<-q_1,-q_4|H_P|-q_4-q_1>-2\omega_1(2\pi)^2\delta^2(q_1-q_4),
\eeq
so that
\[
D_{2a}=i\int\frac{d^2q_1d^2q_4}{(2\pi)^4}
P_{E_{12}}(q_1)P_{E_{34}}(q_4)P_{E_{13}}(q_1)P_{E_{24}}(q_4)<-q_1,-q_4|H_P|-q_4,-q_1>
\]
\beq
+2i\int\frac{d^2q}{(2\pi)^2}\omega(q)
P_{E_{12}}(q)P_{E_{34}}(q)P_{E_{13}}(q)P_{E_{24}}(q)
\label{aa1}
\eeq
and similarly
\[
D_{2b}=i\int\frac{d^2q_1d^2q_4}{(2\pi)^4}
P_{E_{12}}(q_1)P_{E_{34}}(q_4)P_{E_{13}}(q_4)P_{E_{24}}(q_1)<q_1,q_4|H_P|q_4,q_1>
\]
\beq
+2i\int\frac{d^2q}{(2\pi)^2}\omega(q)
P_{E_{12}}(q)P_{E_{34}}(q)P_{E_{13}}(q)P_{E_{24}}(q).
\label{ab1}
\eeq
As we see the additional terms in (\ref{aa1}) and (\ref{ab1})
containing $\omega(q)$
cancel with a similar term in (\ref{ampli1}),
so that the remaining sum of $D_1$ and
$D_2$ turns out to be infrared safe.

So after  cancellation of the gluon Regge trajectories the
infrared safe contributions from
diagrams in Figs. \ref{fig2} and \ref{fig3} are
\beq
D_1=\theta(y)\frac{\partial}{\partial y}
\int_0^ydy'\int \frac{d^2q}{(2\pi)^2}P^2(y-y',q)P^2(y',q)
\label{fina1}
\eeq
and
\beq
D_2=2\theta(y)\int_o^ydy'\int\frac{d^2qd^2q'}{(2\pi)^4}<q,q'|H|q',q>
P(y-y',q)P(y-y',q')P(y',q)P(y',q').
\label{fina2}
\eeq

\section{Two interactions between the projectile and target pomerons}
With two interactions between the target and projectile pomerons
we can use our old results in ~\cite{braun} where coupling of two pomerons
to the BKP 4-gluon case was studied (see also Appendix 2.)

In this case there are transitions both with the redistribution of colour, that is $|(12)(34)>\to|(13)(24)>$, and without this
redistribution, that is  $|(12)(34)>\to|(12)(34)>$.

To continue our line of studies we start with the redistribution
of colour.
As follows from our studies in Appendix 2 in this case between the pomerons
from the projectile and target there can appear two 4-gluon BKP
states $|1243>$ and $|1324>$. Inserting the 4-gluon BKP Green function $G$ between them we find that the projectile and target pomerons will be connected by
\beq
M_E^{(a)}=\frac{1}{4}N_c^2\Big(v_{13}+v_{24}-v_{23}-v_{14}\Big)[G_E^{(1243)}
+G_E^{(1342)}]
\Big(v_{12}+v_{34}-v_{23}-v_{14}\Big),
\eeq
where, say, $G_E^{(1243)}$ is an operator acting in the 4-gluon space
\[<q_1,q_2,q_3,q_4|G_E^{(1243)}|q'_1,q'_2,q'_3,q'_4>\]
satisfying the equation
\beq
(E-H^{(1243)})G_E^{1243}=1
\eeq
with
\beq
H^{(1243)}=-\sum_{i=1}^4\omega_i-\frac{1}{2}g^2N_c(v_{12}+v_{24}+v_{43}+v_{31}).
\eeq
Note that $M^{(a)}_E$ can also be presented in terms of the
infrared safe BFKL Hamiltonian for the pomeron $H_P$
\beq
M_E^{(a)}=\frac{1}{4}\Big(H_{P,13}+H_{P,24}-H_{P,23}-H_{P,14}\Big)[G_E^{(1243)}
+G_E^{(1324)}]
\Big(H_{P,12}+H_{P,34}-H_{P,23}-H_{P,14}\Big)
\eeq
which demonstrates that $M^{(a)}_E$ is infrared safe.

The explicit form for the kernel of $M$ is
\[
<q_1,q_2,q_3,q_4|M_E^{(a)}|q'_1,q'_2,q'_3,q'_4>=2\pi\delta^2\Big(\sum_{j=1}^4q'_j-\sum_{j=1}q_j\Big)
\frac{1}{4}N_c^2
\int\prod_{j=1}^4\frac{d^2k'_j}{(2\pi)^2)}\,\frac{d^2k_j}{(2\pi)^2)}
2\pi\delta^2\Big(\sum_{j=1}^4k'_j-\sum_{j=1}k_j\Big)\]\[
<q_1,q_2,q_3,q_4|v_{13}+v_{24}-v_{23}-v_{14}|k_1,k_2,k_3,k_4>
<k_1,k_2,k_3,k_4|G_E^{(1243)}+G_E^{(1342)}|k'_1,k'_2,k'_3,k'_4>\]\beq
<k'_1,k'_2,k'_3,k'_4|v_{12}+v_{34}-v_{23}-v_{14}|q'_1,q'_2,q'_3,q'_4>,
\label{kernm}
\eeq
where for instance
\beq
<q_1,q_2,q_3,q_4|v_{13}|k_1,k_2,k_3,k_4>=
(2\pi)^4\delta^2(q_2-k_2)\delta^2(q_4-k_4)v(q_1,q_3|k_1,k_3)
\label{v13}
\eeq
and $v(q_1,q_3|k_1,k_3)$ is given by (\ref{eq5}).


The amplitude $D_3$ with the redistribution of colour corresponding to Fig. \ref{fig4}
will be given by
\[D_{3a}=-
\int\frac{dE_1}{2\pi}\frac{dE'_1}{2\pi}
\int\prod_{j=1}^4\frac{d^2q'_j}{(2\pi)^2)}\,\frac{d^2q_j}{(2\pi)^2)}
2\pi\delta^2\Big(\sum_{j=1}^4q'_j-\sum_{j=1}q_j\Big)\]\beq
P_{E-E_1}(q_1)P_{E_1}(q_4)
<q_1,-q_1,-q_4,q_4|M_E^{(a)}|q'_1,-q'_4,-q'_2,q'_4>
P_{E-E'_1}(q'_1)P_{E'_1}(q'_4).
\label{a3a}
\eeq
The prefactor includes two $(-i)$ from $\hat{V}$.

The amplitude without colour redistribution will differ from
$D_{3a}$ in that between the projectile and target pomerons now appear
four BKP states $|1234>,\,|1243>,\,|1342>$
and $|1432>$ and their coupling to the projectile and target
pomerons will be the same. This means that now the projectile and
target pomerons will be connected by
\[
M_E^{(b)}=\frac{1}{4}N_c^2\Big(v_{13}+v_{24}-v_{23}-v_{14}\Big)
[G_E^{(1234)}+G_E^{(1432)}+G_E^{(1243)}+G_E^{(1342)}]
\Big(v_{13}+v_{24}-v_{23}-v_{14}\Big)\]\[ =
\frac{1}{4}\Big(H_{P,13}+H_{P,24}-H_{P,23}-H_{P,14}\Big)
[G_E^{(1234)}+G_E^{(1432)}+G_E^{(1243)}+G_E^{(1342)}]\]\beq
\Big(H_{P,13}+H_{P,24}-H_{P,23}-H_{P,14}\Big).
\eeq
The remaining formulas do not change and we find that
the amplitude $D_{3b}$ without colour  redistribution corresponding to Fig. \ref{fig4}
will be given by
\[D_{3b}=-
\int\frac{dE_1}{2\pi}\frac{dE'_1}{2\pi}
\int\prod_{j=1}^4\frac{d^2q'_j}{(2\pi)^2)}\,\frac{d^2q_j}{(2\pi)^2)}
2\pi\delta^2\Big(\sum_{j=1}^4q'_j-\sum_{j=1}q_j\Big)\]\beq
P_{E-E_1}(q_1)P_{E_1}(q_4)
<q_1,-q_1,-q_4,q_4|M_E^{(b)}|q'_1,-q'_1,-q'_4,q'_4>
P_{E-E'_1}(q'_1)P_{E'_1}(q'_4).
\label{a3b}
\eeq

To pass to rapidity, consider first evolution of the momentum integrand in time.
\beq
Z_1=\int \frac{dEdE_1dE_2}{(2\pi)^3}e^{-iEt}P_{E-E_1}P_{E_1}P_{E-E_2}
P_{E_2}M(E),
\eeq
where we suppress the obvious momentum dependence.
Presenting
\[P_E=\int dtP(t)e^{iEt},\ \ M(E)=\int d\tau M(\tau)e^{iE\tau}\]
we have
\[
Z_1(t)=\int dt_1dt_2dt_3dt_4d\tau\int\frac{dEdE_1dE_2}{(2\pi)^3}
e^{i(E-E_1)t_1+iE_1t_2+i(E-E_2)t_3+iE_2t_4+iE\tau}\]\beq
P_1(t-t_1)P_2(t-t_2)P_3(t_3)P_4(t_4)M(\tau)=
\int_0^t dt_1\int_0^{t_1}dt_2 P_1(t-t_1)P_2(t-t_1)M(t_1-t_2)P_3(t_2)P_4(t_2)
,\eeq
where we take into account that both $P(t)$ and $M(t)$ are zero at $t<0$.
Analytically continuing to rapidities we find
\beq
Z_1(y)=-\int_0^y dy_1\int_0^{y_1} P_1(y-y_1)P_2(y-y_2)M(y_1-y_2)P_3(y_2)
P_4(y_2).
\eeq

So in the end
\[
D_{3a}=\theta(y)\int_0^y dy_1\int_0^{y_1}dy_2
\int\prod_{j=1}^4\frac{d^2q'_j}{(2\pi)^2)}\,\frac{d^2q_j}{(2\pi)^2)}
2\pi\delta^2\Big(\sum_{j=1}^4q'_j-\sum_{j=1}q_j\Big)\]\beq
P_{12}(y-y_1,q_1)P_{34}(y-y_1,q_4)
<q_1,-q_1,-q_4,q_4|M^{(a)}(y_1-y_2)|q'_1,-q'_4,-q'_2,q'_4>
P_{13}(y_2,q'_1)P_{24}(y_2,q'_4)
\label{a3ta}
\eeq
and
\[
D_{3b}=\theta(y)\int_0^y dy_1\int_0^{y_1}dy_2
\int\prod_{j=1}^4\frac{d^2q'_j}{(2\pi)^2)}\,\frac{d^2q_j}{(2\pi)^2)}
2\pi\delta^2\Big(\sum_{j=1}^4q'_j-\sum_{j=1}q_j\Big)\]\beq
P_{12}(y-y_1,q_1)P_{34}(y-y_1,q_4)
<q_1,-q_1,-q_4,q_4|M^{(b)}(y_1-y_2)|q'_1,-q'_4,-q'_2,q'_4>
P_{12}(y_2,q'_1)P_{34}(y_2,q'_4),
\label{a3tb}
\eeq
where we indicated the numbers of reggeized gluons of which  different
pomerons are made.

\section{The deuteron-deuteron scattering}
The total deuteron-deuteron scattering cross-section apart from the contributions studied in the
previous sections will include the standard single and double scattering
terms (see Appendix 1. Figs.
\ref{fig2ap} and \ref{fig6ap}). So the total cross-section is the sum
\beq
\sigma^{dd}=\sigma^{single}+\sigma^{double}+\sum_{i=1}^3\sigma^{(i)}.
\label{sigddd}
\eeq
Here the single cross-section is well known
\beq
\sigma^{single}=\sigma_{pp}+\sigma_{nn}+2\sigma_{pn}.
\eeq
The double cross-section is (see Appendix 1.)
\beq
\sigma^{double}=-\frac{1}{2}(\sigma_{pp}\sigma_{nn}+\sigma_{pn}^2)
\int d^2bT^2_d(b),
\eeq
where the transverse density $T(b)$ is expressed in the standard manner via the deuteron wave function:
\beq
T_d(b)=\int dz|\psi_d(b,z)|^2.
\eeq
Finally the additional cross-sections due to the QCD effects are
expressed via $D_i$, $i=1,2,3$ according to Eq. (\ref{sigdd})
\beq
\sigma^{(i)}=-2D_iN_c^2\Big(<\frac{1}{2\pi r^2}>_d\Big)^2,
\eeq
where $<...>_d$ means averaging in the deuteron.

\section{Discussion}
For high-energy nucleus-nucleus scattering we have calculated the leading
terms in the eikonal function
for the forward scattering amplitude corresponding to the collision of
two scattering
centers in the projectile nucleus with two scattering centers in the
target nucleus.
Apart from the obvious pomeron exchange, these terms include
contributions from connected diagrams involving two pomerons from the
projectile and two pomerons from the target with all possible interactions in between.
We have demonstrated that the result is infrared safe, as one expected.
It is remarkable that it involves contributions from intermediate BKP states
formed by 4 reggeized gluons.

For the deuteron-deuteron scattering the total cross-section is the sum of the eikonal function
plus the double scattering term in Eq. (\ref{sigddd}). If one takes the
pomerons as
described by the BFKL equation then the behavour of the $dd$ cross-section will roughly correspond to the
double pomeron exchange  that is $~\exp (2\Delta_{BFKL}y)$,
where $\Delta_{BFKL}=(\alpha_sN_c/\pi)4\ln 2$ is the BFKL intercept. Note
that in this
case the intermediate BKP state in $\sigma^{(3)}$ will enter at
comparatively low energies,
so that one cannot use its asymptotical behaviour to find the result.
On the other hand description of the total NN cross-section by the
single pomeron exchange is
obviously unrealistic. The single and double cross-sections in (\ref{sigdd})
can be calculated
using the experimental values of this cross-section. As to the rest, instead
of simple
BFKL pomerons  one may use unitarized expressions corresponding to sums of
fan diagrams and
found as solutions of the BK equation. The latter do not grow at high energies
 and so
the asymptotical behavour of the cross-section  will
be determined by that of the BKP state entering $\sigma^{(3)}$,
which grows, although not so fast as the
pomeron ($\sim\exp 0.243\Delta_{BFKL}y$, \cite{korch}). As a result the $dd$ cross-sections allow for the direct
experimental study of the behaviour of
such states, theoretical  calculation of which presents serious difficulties.

This is of course true also for collisions of heavy nuclei. But in this case
unitarization of the
eikonal function will in any case lead to the total cross-section
which are more or less geometrical.
Additional terms in the eikonal may of course change it considerably
but the cross-section will not
change at least inside the nucleus where the eikonal remains large.
In this case a more interesting problem is the inclusive gluon
production in the nucleus-nucleus collisions to which disconnected
diagrams do not contribute. This requires cutting our
forward scattering
amplitude to select the observed intermediate states with due attention to
possible
cancellation between real and virtual processes (AGK cancellations).
This problem will be
dealt in future studies.

\section{Acknowledgments}
This work has been supported by grant  RFFI 12-02-00356-a.
The author is thankful to J.B.Bartels and G.P.Vacca for their interest in this study and
helpful discussions. He also thanks the INFN and Universities of Bologna
and Hamburg for hospitatlty.

\section{Appendix 1. Deuteron in the Glauber approach}
\subsection{Scattering on the deutron}
\begin{figure}
\begin{center}
\includegraphics[scale=0.85]{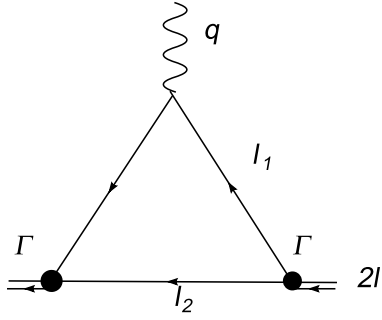}
\end{center}
\caption{Electromagnetic form factor of the deuteron}
\label{fig1ap}
\end{figure}

To formulate the Glauber approach to the collisions with deuteron
in the diagrammatic technique we first have to relate the relativistic
$dpn$ vertex $\Gamma$ with the deuteron wave function. To this end we
study the electromagnetic form-factor of the deuteron, illustrated in
Fig. \ref{fig1ap}, where vertices $\Gamma$ are shown with blobs.
In the lab. sysstem and at zero momentum transfer it is equal
to $2M$ where $M=2m-\epsilon$ is the deuteron mass. So we get the
normalization condition
\beq
\int \frac{d^4l_2}{((2\pi)^4 i}\frac {2l_{10}\Gamma^2({\bf l}_2)}
{(m^2-l_2^2-i0)(m^2-l_1^2-i0)^2}=2M.
\label{norm}
\eeq
Here $l_1+l_2=2l$ and $4l^2=M^2$. We neglect spins and
consider all particles as scalar for simplicity.
We have $l_0=m-\epsilon/2$, and ${\bf l}=0$, so that putting $l_2=
l+\lambda$ we find
\[m^2-l_2^2=-2m\lambda_0-\lambda_\perp^2,\ \
m^2-l_1^2=2m\lambda_0-\lambda_\perp^2,\]
where we used the orders of magnitude
$\lambda_0\sim\epsilon$, $|\lambda_\perp|\sim\sqrt{m\epsilon}$.
Integration over $\lambda_0$ transforms (\ref{norm}) into
\beq
\int \frac{d^3l_2}{(2\pi)^3} \frac {\Gamma^2({\bf l}_2)}
{8m(m\epsilon+{\bf l}_2^2)^2}=2.
\label{norm1}
\eeq
Comparing with the standard normalization of the
deuteron wave function $\psi_d({\bf l})$ we find the desired
relation
\beq
\psi_d({\bf l})=\frac{\Gamma({\bf l})}
{4(2\pi)^{3/2}\sqrt{m}(m\epsilon+{\bf l}^2)},
\label{psid}
\eeq
which allows to relate the relativistic $dpn$ vertex with the
deuteron wave function in the momentum space.

\begin{figure}
\begin{center}
\includegraphics[scale=0.85]{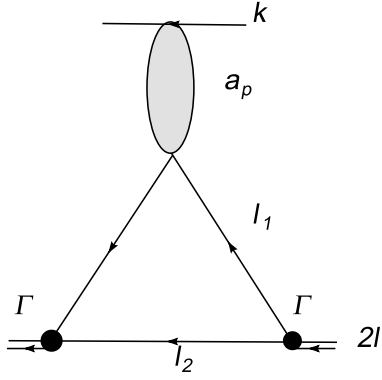}
\end{center}
\caption{Scattering on the deuteron in the impulse approximation}
\label{fig2ap}
\end{figure}

In the
impulse approximation, Fig. \ref{fig2ap}, with the spectator neutron
the corresponding amplitude is given by
\beq
{\cal A}^{imp}=a\int \frac{d^4l_2}{(2\pi)^4 i}
\frac {\Gamma^2({\bf l}_2)}
{(m^2-l_2^2-i0)(m^2-l_1^2-i0)^2},
\eeq
where $a$ is the forward scattering amplitude on the proton.
Using (\ref{norm}) we find that the integral is equal to 2 and
we get
$
{\cal A}^{imp}=2a.
$
But the relativistic flux on the deuteron is twice that on the proton,
so that we get
$\sigma_d=\sigma_p+\sigma_n$,
where the second term takes into account the diagram of Fig.
\ref{fig2ap} with the spectator proton.

\begin{figure}
\begin{center}
\includegraphics[scale=0.85]{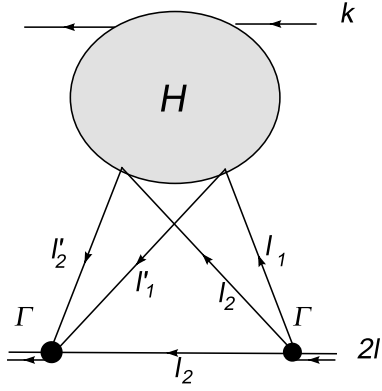}
\end{center}
\caption{Double scattering on the deuteron}
\label{fig3ap}
\end{figure}

Now consider double scattering on the deuteron illustrated in Fig.
\ref{fig3ap}.
The amplitude is given by
\beq
{\cal A}=\int\frac{d^4l_2}{(2\pi)^4 i}\frac{d^4l'_2}{(2\pi)^4 i}
H(l_{2z}-l'_{2z})
\frac{\Gamma({\bf l}_2)}{(m^2-l_1^2-i0)(m^2-l_2^2-i0)}
\frac{\Gamma({\bf l'}_2)}{(m^2-{l'_1}^2-i0)(m^2-{l'_2}^2-i0)},
\label{double}
\eeq
where $H$ is the high-energy part and it is taken into account that
it can only depend on the $z$-component of the transferred momentum,
since it is the only of the spatial components which enters multiplied
by the high projectile momentum.

Integrations over the zero components of the nuclear momenta factorize
and we get
\beq
{\cal A}=\int\frac{d^3l_2d^3l'_2}{(2\pi)^3}
H(l_{2z}-l'_{2z})
\frac{\Gamma({\bf l}_2)}{4m(m\epsilon+{\bf l}_2^2)}
\frac{\Gamma({\bf l'}_2)}{4m(m\epsilon+{{\bf l'}_2}^2)},
\label{double1}
\eeq
or using (\ref{psid})
\beq
{\cal A}=\frac{1}{m}\int\frac{d^3l_2d^3l'_2}{(2\pi)^3 }
H(l_{2z}-l'_{2z})\psi_d({\bf l}_2)\psi_d({\bf l'}_2).
\label{double2}
\eeq
Integrations over the transverse components are done immediately to
convert the wave functions into those with the transverse coordinates zero:
\beq
{\cal A}=\frac{1}{m}\int\frac{dl_{2z}dl'_{2z}}{2\pi }
H(l_{2z}-l'_{2z})\psi_d(r_\perp=0,l_{2z})\psi_d(r_\perp=0, l'_{2z}).
\label{double3}
\eeq
Transforming completely to the coordinate space we find
\beq
{\cal A}=\frac{1}{m}\int\frac{dl_{2z}}{2\pi}\frac{dq_{z}}{2\pi }
dzdz'
H(q_z)\psi_d(r_\perp=0,z)\psi_d(r_\perp=0,z')
e^{il_{2z}(z-z')-iq_zz'},
\label{double4}
\eeq
where we introduced the $z$-component of the transferred momentum
putting $l'_z=l_z+q_z$.
Integration over $l_{2z}$ gives our final expression
\beq
{\cal A}=\frac{1}{m}\int\frac{dq_{z}}{2\pi }
dz
H(q_z)|\psi_d(r_\perp=0,z)|^2
e^{-iq_zz}.
\label{double5}
\eeq

The Glauber approximation follows if $H(q_z)$ has a singularity
at $q_z=0$. Typically
\beq
{\rm Im}\,H(q_z)=-\hat{D}(2\pi)\delta(2kq)=-\hat{D}\frac{\pi}{k_0}\delta(q_z).
\label{handf}
\eeq
Here we use $q_0<<|q_z|$ and $k_0=k_z>0$.
In this case we get the Glauber approximation for the double scattering
amplitude
\beq
{\rm Im}\,{\cal A}=-\frac{1}{s}\hat{D}
\int dz|\psi_d(r_\perp=0,z)|^2=-\frac{1}{ s}\hat{D}<\frac{1}{2\pi r^2}>_d.
\label{double6}
\eeq
where $<...>_d$ means the average in the deuteron. The cross-section is
\beq
\sigma_d=-\frac{1}{2s^2}\hat{D}<\frac{1}{2\pi r^2}>_d.
\label{sid}
\eeq

\begin{figure}
\begin{center}
\includegraphics[scale=0.65]{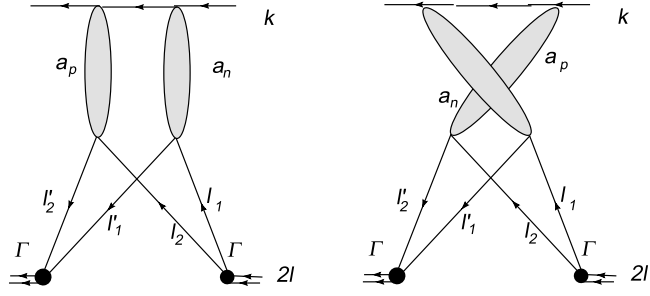}
\end{center}
\caption{Double elastic collision on the deuteron}
\label{fig4ap}
\end{figure}

To see how this formula works consider the simplest case of the
double scattering corresponding to double elastic collision
shown in Fig. \ref{fig4ap}.
In this case
\beq
H(q_z)=ia_pa_n\Big(\frac{-i}{m_2-(k+q_z)^2-i0}+
\frac{-i}{m_2-(k-q_z)^2-i0}\Big)=ia_pa_n 2\pi \delta(2k_0q_z),
\label{hsimple}
\eeq
so that
$\hat{D}=-a_pa_n$.
Using (\ref{double6}) we find
\beq
{\rm Im}{\cal A}=a_pa_n
\frac{1}{s} <\frac{1}{2\pi r^2}>_d=
-s\sigma_p\sigma_n<\frac{1}{2\pi r^2}>_d.
\label{simple1}
\eeq
From this dividing by $2s$ and doubling to take into account the crossed
diagram find the double cross-section
\beq
\sigma^{double}=-\sigma_p\sigma_n<\frac{1}{2\pi r^2}>_d.
\label{simple2}
\eeq

For the nuclear target instead of (\ref{sid})
we have at fixed impact parameter $b$
\beq
\sigma_A(b)=-\frac{1}{2s^2}\hat{D}T_A^2(b)
\label{sia}
\eeq
and for double elastic collisions instead of (\ref{simple2})
\beq
\sigma_A(b)=-\frac{1}{2}A(A-1)\sigma_N^2T_A^2(b).
\label{simplea}
\eeq

To conclude we note that $H$ and $F$ are  both Lorenz invariant.
So (\ref{handf}) can be used to find $F$ in any system.

\subsection{Double scattering in  d-d collisions}
\begin{figure}
\begin{center}
\includegraphics[scale=0.65]{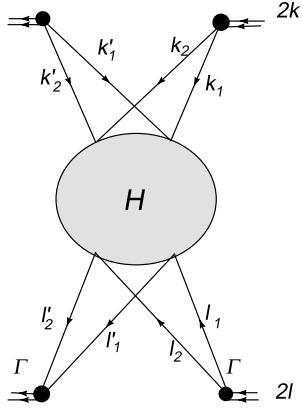}
\end{center}
\caption{Double scattering in deuteron-deuteron collisions}
\label{fig5ap}
\end{figure}

Now consider the case when two deuterons collide at high energies
and each one experiences double collision, illusttated in
Fig. \ref{fig5ap}. Our treatment is to consider subsequently the
two systems in which the deuteron is well understandable, the rest
systems of the target and projectile deuterons.

We start from the rest system of the target deuteron. We use Eq.
(\ref{double6}) and write
\beq
{\rm Im}\,{\cal A}=-\frac{1}{ s}\hat{D}_1<\frac{1}{2\pi r^2}>_d,
\eeq
where it is assumed that ${\rm Im}\, H_1=-\hat{D}_12\pi\delta(2kq)$ and we
include into $H_1$
all the rest part of the diagram in Fig. \ref{fig5ap} including
the coupling to the projectile nucleons.

Now we  boost  the system into the rest one for
the projectile. In this system
 we can repeat our treatment
of the coupling to the two nucleons. If
$\hat{D}_1=\hat{D}2\pi\delta(2l\kappa)$ where $\kappa$ is the momentum
transferred from the projectile, integration over the nucleon momenta
will give the same factor
$ (1/s)<1/2\pi r^2>_d$
and we shall get
\beq
{\rm Im}\,{\cal A}=-\Big(\frac{1}{s})^2\hat{D}<\frac{1}{2\pi r^2}>^2_d.
\label{double8}
\eeq
Note that we have
\beq
{\rm Im}\,H=-\hat{D}(2\pi)^2\delta(2(l\kappa))\delta(2(kq)).
\label{hdd}
\eeq
This means that
(\ref{double8}) is symmetric
in projectile and target, as expected.
From (\ref{double8}) we immediately get the cross-section (\ref{sigdd})
taking in to account the definition of $D$, Eq. (\ref{ddef})

\begin{figure}
\begin{center}
\includegraphics[scale=0.65]{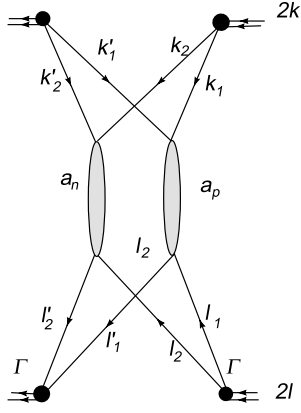}
\end{center}
\caption{Double elastic collision in deuteron-deuteron scattering}
\label{fig6ap}
\end{figure}

Special attention is to be given for the case when the
high-energy part $H$ is disconnected, shown in Fig. \ref{fig6ap}.
Then $H$ contains a $\delta$ function corresponding to conservation
laws for the two connected parts and is given by
\beq
H=i(a_{pp}a_{nn}+a_{pn}^2)(2\pi)^4\delta^4(k_1+l_1-k'_1-l'_1).
\eeq
Factor $i$ combines $(-i)$ from the definition of ${\cal A}$ and
$i^2$ from the two NN amplitudes.
Note that
\[(2\pi)^4\delta^4(k_1+l_1-k'_1-l'_1)=(2\pi)^4\delta(\kappa_++q_+)
\delta(\kappa_+q_-)
\delta^2(k_{1\perp}+l_{1\perp}-k'_{1\perp}1-l'_{1\perp})\]\[
=2s(2\pi)^2\delta(2(l\kappa))\delta(2(kq))
\int d^2be^{ib(k_{1\perp}+l_{1\perp}-k'_{1\perp}1-l'_{1\perp})}.\]
As we see, integrations over the transverse coordinates of the
nucleons in the projectile and target become interdependent.
Under the sign of integration over $b$ we include the exponentials
depending on the transverse momenta of the projectile and target
in the corresponding integrals to obtain in (\ref{double3})
$\psi_d(b,l_{2z})\psi_d(b, l'_{2z})$ instead of
$\psi_d(r_\perp=0,l_{2z})\psi_d(r_\perp=0, l'_{2z})$
and similarly for the projectile. All subsequent calculations remain
unchanged and in the end we obtain in (\ref{double6})
\beq
\int dz|\psi_d(b,z)|^2\equiv T_d(b)
\label{tdb}
\eeq
instead of
\[\int dz|\psi_d(r_\perp=0,z)|^2=<\frac{1}{2\pi r^2}>_d.\]
So the net result of the connection between the
transferred transverse momenta is to substitute
$
<1/2\pi r^2>_d\to T_d(b)
$
 both in the projectile and target and then integrate over $b$.
The rest factors from $H$ give $\hat{D}=2sa_pa_n$ and from
(\ref{double8}) we conclude
\beq
{\rm Im}\,{\cal A}=\frac{2}{s}(a_{pp}a_{nn}+a_{pn}^2)\int d^2bT^2_d(b)
\eeq
Dividing by $4s$ we find the cross-section
\beq
\sigma_dd=-\frac{1}{2}(\sigma_{pp}\sigma_{nn}+\sigma_{pn}^2)
\int d^2bT^2_d(b)
\label{double9}
\eeq
which looks very much like the standard Glauber formula.

For the collision of two heavy nuclei instead of Eq. (\ref{double9})
we shall get at fixed $b$
\beq
\sigma_{AB}(b)=-\frac{1}{4}A(A-1)B(B-1)\sigma_N^2
\Big(\int d^2b_AT_A(b_A)T_B(b-b_A)\Big)^2.
\label{double10}
\eeq

\section{Appendix 2. Colour factors}

The explicit expressions for the colour wave functions of the projectile and
target
 in which pairs (12),(34) and (13),(24) respectively form  colour singlets
are:
\beq
|(12)(34)>=\frac{1}{N_c^2}\delta_{a_1a_2}\delta_{a_3a_4},\ \
|(13)(24)>=\frac{1}{N_c^2}\delta_{a_1a_3}\delta_{a_2a_4}.
\eeq
Here we neglect terms of the relative order $1/N_c^2$.
Their scalar product is
\beq
<(13)(24)|(12)(34)>=\frac{1}{N_c^4}\delta_{a_1a_3}\delta_{a_2a_4}\delta_{a_1a_2}\delta_{a_3a_4}
=\frac{1}{N_c^2},
\label{redisfac}
\eeq
which is the overall damping factor accompanying all diagrams with the redistribution
of colour like Fig. \ref{fig1}.

We denote $C_{ij}=-(T_iT_j)$. For interactions connecting vacuum pairs either in the projectile or in
the target $C_{ij}=N_c$
\beq
<(13)(24)|C_{ij}|(12)(34)>=N_c<(13)(24)|(12)(34)>=\frac{1}{N_c},\ \
(ij)=(12),\  (34),\ (13),\ (24)
\eeq
For the remaining two interactions we find
\beq
<(13)(24)|C_{14}|(12)(34)>=\frac{1}{N_c^4}
\delta_{a'_1a'_3}\delta_{a'_2a'_4}\delta_{a_1a_2}\delta_{a_3a_4}
\delta_{a'_2a_2}\delta_{a'_3a_3}f^{a'_1a_1c}f^{a'_4a_4c}=
\frac{1}{N_c^4}
f^{a_3a_1c}f^{a_1a_3c}=-\frac{1}{N_c}.
\label{c14}
\eeq
Interchange $(1\lra 2),(3\lra 4)$ gives
\beq
<(13)(24)|C_{23}|(12)(34)>=-\frac{1}{N_c}.
\label{c23}
\eeq
So effectively for these interactions $C_{ij}=-N_c$

Apart from states $|(12)(34)>$ and $|(13)(24)>$ in the diagrams we  encounter
six BKP states with different ordering of the 4 gluons:
\[|1234>,\ |1243>,\ |1324>,\ 1342>,\ |1423>,\ 1432>.\]
In the high colour limit their explicit form is
\beq
|1234>=\frac{1}{2N_c^2}h^{a_1a_2c}h^{a_3a_4c},
\label{bkp}
\eeq
where
$
h^{abc}=d^{abc}+if^{abc}
$ with the properties
\beq
[h^{abc}]^*=h^{bac},\ \sum_{cd}[h^{acd}]^*h^{bcd}=
\delta_{ab}2N_c\Big(1-\frac{2}{N_c^2}\Big),\ \
\sum_{cd}h^{acd}h_{bcd}=-\delta_{ab}\frac{4}{N_c}
\eeq
The states $|ijkl>$ are cyclic symmetric in $(ijkl)$.

Their scalar products with the projectile and target states are
\beq
<(12)(34)|1234>=\frac{1}{N_c^4}\delta_{a_1a_2}\delta_{a_3a_4}
h_{a_1a_2c}h_{a_3a_4c}=0.
\eeq
\beq
<(13)(24)|1234>=\frac{1}{N_c^4}\delta_{a_1a_3}\delta_{a_2a_4}
h_{a_1a_2c}h_{a_3a_4c}=\frac{1}{N_c^4}h^{a_1a_2c}h^{a_1a_4c}=-2\frac{1}{N_c^3}.
\label{nonort}
\eeq
Generally if (12) or (34) are neighbors in (ijkl) then states $|(12)(kl)>$ and $|ijkl>$
are orthogonal. If they are not the the scalar product is the same as in (\ref{nonort}).

We also need matrix elements of colour matrices $C_{ij}$ between projectile
(target)  states and BKP states. Obviously we need only $C_{ij}$ which do not
connect vacuum pairs in the projectile (target), namely for $(ij)=(13),(14),(23),(24)$
Then  we find that for $(klmn)=((1234)$, $(2134)$, $(2143)$ and $(1423)$
\beq
<(12)(34)|C_{ij}|klmn>=\pm\frac{1}{2},
\label{cij}
\eeq
where the sign plus is to be taken when  $(ij)$ are neighbours in $(klmn)$
and the sign minus when they are not.
Acting on  the rest two states  $|1324>$ and $|1423>$.
all matrices $C_{13}, C_{24}, C_{14}$ and $C_{23}$ give $N_c/2$
since neighbour gluons are in the gluon colour state and the matrix elements
become damped by $1/N_c^2$. E.g.
\beq
<(12)(34)|C_{13}|1423>=\frac{N_c}{2}<(12)(34)|1423>=-\frac{1}{N_c^2}.
\label{cij1}
\eeq
(Note that the correct derivation of (\ref{cij}) and \ref{cij1})
in some cases requires taking into account subdominant terms in (\ref{bkp}))

As a result the matrix elements of $C_{13}$ are
\[<(12)(34)|C_{13}|1234>=<(12)(34)|C_{13}|2143>=-\frac{1}{2},\]
\[<(12)(34)|C_{13}|2134>=<(12)(34)|C_{13}|1243>=\frac{1}{2},\]
\[<(12)(34)|C_{13}|1324>=<(12)(34)|C_{13}|1423>=0,
\]
which implies
\beq
<(12)(34)|C_{13}=-\frac{1}{2}<1234|-\frac{1}{2}<2143|+\frac{1}{2}
<2134|+\frac{1}{2}<1243|.
\eeq
Note that the summed probabilities  correctly give unity.

Similarly
\[
<(12)(34)|C_{24}=-\frac{1}{2}<1234|-\frac{1}{2}<2143|+\frac{1}{2}
<2134|+\frac{1}{2}<1243|,
\]
\[
<(12)(34)|C_{23}=-\frac{1}{2}<2134|-\frac{1}{2}<1243|+\frac{1}{2}
<1234|+\frac{1}{2}<2143|,
\]
and
\[
<(12)(34)|C_{14}=-\frac{1}{2}<2134|-\frac{1}{2}<1243|+\frac{1}{2}
<1234|+\frac{1}{2}<2143|.
\]

Interchanging  $(2\lra 3)$ we get
\[
C_{12}|(13)(24)>=-\frac{1}{2}|1324>-\frac{1}{2}|3142>+\frac{1}{2}
|3124>+\frac{1}{2}|1342>,
\]\[
C_{34}|(13)(24)>=-\frac{1}{2}|1324>-\frac{1}{2}|3142>+\frac{1}{2}
|3124>+\frac{1}{2}|1342>,
\]\[
C_{23}|(13)(24)>=-\frac{1}{2}|3124>-\frac{1}{2}|1342>+\frac{1}{2}
|1324>+\frac{1}{2}|3142>,
\]\[
C_{14}|(13)(24)>=-\frac{1}{2}|1324>-\frac{1}{2}|3142>+\frac{1}{2}
|3124>+\frac{1}{2}|1342>.
\]

These relations allow to study matrix elements of the product of two
matrices $C_{ij}C_{kl}$ between the projectile and target states.
They are shown in  Table 1. with lines $(ij)$ and columns $(kl)$

\begin{table}
\begin{center}
\caption{Matrix elements of the product $C_{ij}$ (lines) by $C_{kl}$ (columns)
between states $<(12)(34)|$ and $|(13)(24)>$}
\medskip
\begin{tabular}{|c|c|c|c|c|}
\hline
  &(12)&(34)&(23)&(14)\\\hline
(13)&1/2&0&$-1/2$&$-1/2$\\\hline
(24)&0&$-1/2$&0&0\\\hline
(23)&$-1/2$&0&1/2&1/2\\\hline
(14)&$-1/2$&0&1/2&1/2\\\hline
\end{tabular}
\end{center}
\end{table}

From these results we can find the probability to find a particular BKP state
between the projectile and target. States $|1234>$ and $|1324>$ do not appear
and we find the contribution from the double interaction
$V_{ij}V_{kl}$ where $V_{ij}=-C_{ij}g^2v_{ij}$
\beq
\frac{1}{4}<(12)(34)|\Big(v_{13}+v_{24}-v_{23}-v_{14}\Big)
\Big(|1243><1243+|1342><1342|\Big)
\Big(v_{12}+v_{34}-v_{23}-v_{14}\Big)|(13)(24)>.
\label{prob1}
\eeq

We are also  interested in the matrix elements
of products of two matrices $C_{ij}C_{kl}$ between projectile and target states without redistribution of color, that is between states $<(12)(34)|$ and $|(12)(34)>$. In particular we
shall be interested in separate contribution from BKP states.
In this case four different BKP states appear between the
projectile and target $|1234>,\, |1432>,\, |1342>$ and $|1243>$
with equal probability and, similar to (\ref{prob1}) we find
the probabilities
\[
\frac{1}{4}<(12)(34)|\Big(v_{13}+v_{24}-v_{23}-v_{14}\Big)\]\beq
\Big(|1234><1234|+|1243><1243+|1432><1432|+|1342><1342|\Big)
\Big(v_{13}+v_{24}-v_{23}-v_{14}\Big)|(13)(24)>.
\label{prob2}
\eeq


\end{document}